\documentclass[prl,reprint,showpacs,showkeys,showpacs,times,superscriptaddress]{revtex4-1}
\usepackage{amsmath,amssymb,amsthm,times,graphics,graphicx,bm,subfigure}

\usepackage[colorlinks,linkcolor=red,citecolor=blue,urlcolor=blue]{hyperref}

\newcommand{\be}{\begin{equation}}
\newcommand{\ee}{\end{equation}}
\newcommand{\ben}{\begin{eqnarray}}
\newcommand{\een}{\end{eqnarray}}
\newcommand{\bes}{\begin{subequations}}
\newcommand{\ees}{\end{subequations}}
\newcommand{\bF}{\begin{figure}}
\newcommand{\eF}{\end{figure}}

\def\ket#1{ | #1 \rangle}

\def\pd2v#1#2#3{\frac{\partial^2 #1}{\partial #2 \partial #3}}

\def \2x2mat#1#2#3#4{
\left( \begin{array}{cc}
#1 &  #2 \\  #3 &  #4
\end{array} \right)
}

\begin{document}
\title{Experimental nonlocality-based network diagnostics of mutipartite entangled states}

\author{Mario A. Ciampini}\email{marioarnolfo.ciampini@uniroma1.it}
\affiliation{Dipartimento di Fisica, Sapienza Universit\`a di Roma, P.le Aldo Moro 5, 00185, Rome, Italy}

\author{Caterina Vigliar}
\affiliation{Dipartimento di Fisica, Sapienza Universit\`a di Roma, P.le Aldo Moro 5, 00185, Rome, Italy}

\author{Valeria Cimini}
\affiliation{Dipartimento di Fisica, Sapienza Universit\`a di Roma, P.le Aldo Moro 5, 00185, Rome, Italy}

\author{Stefano Paesani}
\affiliation{Dipartimento di Fisica, Sapienza Universit\`a di Roma, P.le Aldo Moro 5, 00185, Rome, Italy}
\affiliation{Centre for Quantum Photonics, H. H. Wills Physics Laboratory and Department of Electrical and Electronic Engineering, University of Bristol, Merchant Venturers Building, Woodland Road, Bristol BS8 1UB, UK.}

\author{Fabio Sciarrino}
\affiliation{Dipartimento di Fisica, Sapienza Universit\`a di Roma, P.le Aldo Moro 5, 00185, Rome, Italy}

\author{Andrea Crespi}
\affiliation{Istituto di Fotonica e Nanotecnologie - Consiglio Nazionale delle Ricerche
(IFN-CNR), P.za Leonardo da Vinci, 32, I-20133 Milano (MI), Italy}
\affiliation{Dipartimento di Fisica - Politecnico di Milano, P.za Leonardo da Vinci, 32, I-20133 Milano (MI), Italy}

\author{Giacomo Corrielli}
\affiliation{Istituto di Fotonica e Nanotecnologie - Consiglio Nazionale delle Ricerche
(IFN-CNR), P.za Leonardo da Vinci, 32, I-20133 Milano (MI), Italy}
\affiliation{Dipartimento di Fisica - Politecnico di Milano, P.za Leonardo da Vinci, 32, I-20133 Milano (MI), Italy}

\author{Roberto Osellame}
\affiliation{Istituto di Fotonica e Nanotecnologie - Consiglio Nazionale delle Ricerche
(IFN-CNR), P.za Leonardo da Vinci, 32, I-20133 Milano (MI), Italy}
\affiliation{Dipartimento di Fisica - Politecnico di Milano, P.za Leonardo da Vinci, 32, I-20133 Milano (MI), Italy}

\author{Paolo Mataloni}
\affiliation{Dipartimento di Fisica, Sapienza Universit\`a di Roma, P.le Aldo Moro 5, 00185, Rome, Italy}

\author{Mauro Paternostro} 
\affiliation{Centre for Theoretical Atomic, Molecular and Optical Physics, School of Mathematics and Physics, Queen's University Belfast, Belfast BT7 1NN, United Kingdom}

\author{Marco Barbieri} 
\affiliation{Dipartimento di Scienze, Universit\`a degli Studi Roma Tre, Via della Vasca Navale 84, 00146, Rome, Italy}

\begin{abstract}
Quantum networks of growing complexity play a key role as resources for quantum computation; the ability to identify the quality of their internal correlations will play a crucial role in addressing the buiding stage of such states.
We introduce a novel diagnostic scheme for multipartite networks of entangled particles, aimed at assessing the quality of the gates used for the engineering of their state. Using the information gathered from a set of suitably chosen multiparticle Bell tests, we identify conditions bounding the quality of the entangled bonds among the elements of a register. We demonstrate the effectiveness, flexibility, and diagnostic power of the proposed methodology by characterizing a quantum resource engineered combining two-photon hyperentanglement and photonic-chip technology.
Our approach is feasible for medium-sized networks due to the intrinsically modular nature of cluster states, and paves the way to section-by-section analysis of large photonics resources.
\end{abstract}

\maketitle

One of the core achievements of recent efforts aimed at the development of quantum technologies is the engineering of ever-larger networks of interacting systems. Quantum networks will play key roles in any embodiment of the upcoming quantum devices, either as distributed architectures for information processing that are naturally able to cope successfully with the detrimental effects of noise~\cite{Kimble08}, or versatile platforms for the quantum simulation of complex processes and dynamics~\cite{Georgescu}. In fact, even devices that are typically conceived and considered as single, monolithic blocks, such as sensors or detectors, actually incorporate highly interconnected units, each with specialised tasks to perform. Such considerations have recently led to the proposal and demonstration of schemes for distributed quantum computing~\cite{distributed,Barz12}, sensing~\cite{Walmsley15}, and cryptography~\cite{Peev09}.  

 %%%%%% Figure 1 %%%%%%
\begin{figure*}[t]
\includegraphics[width=0.811 \columnwidth]{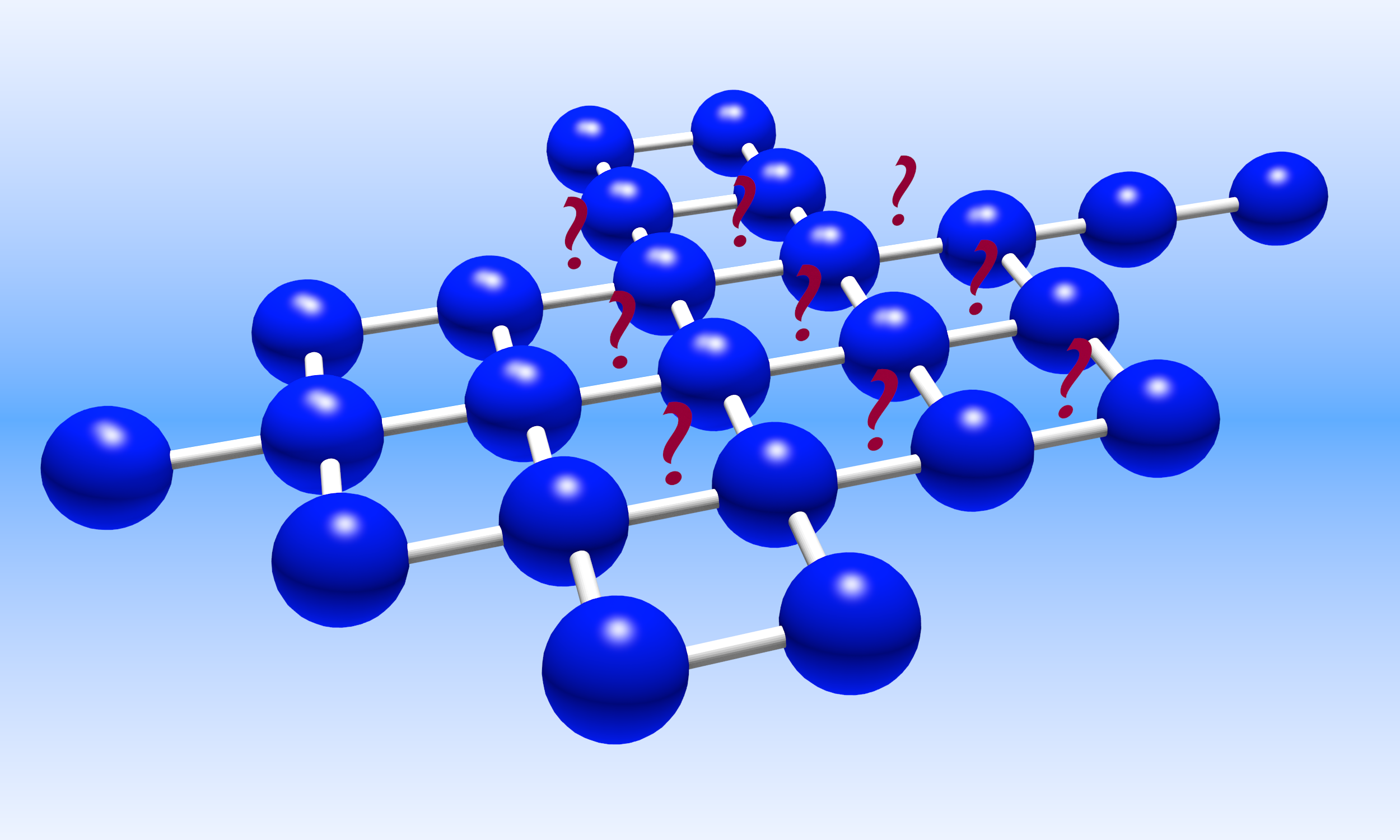}\includegraphics[width=1.2 \columnwidth]{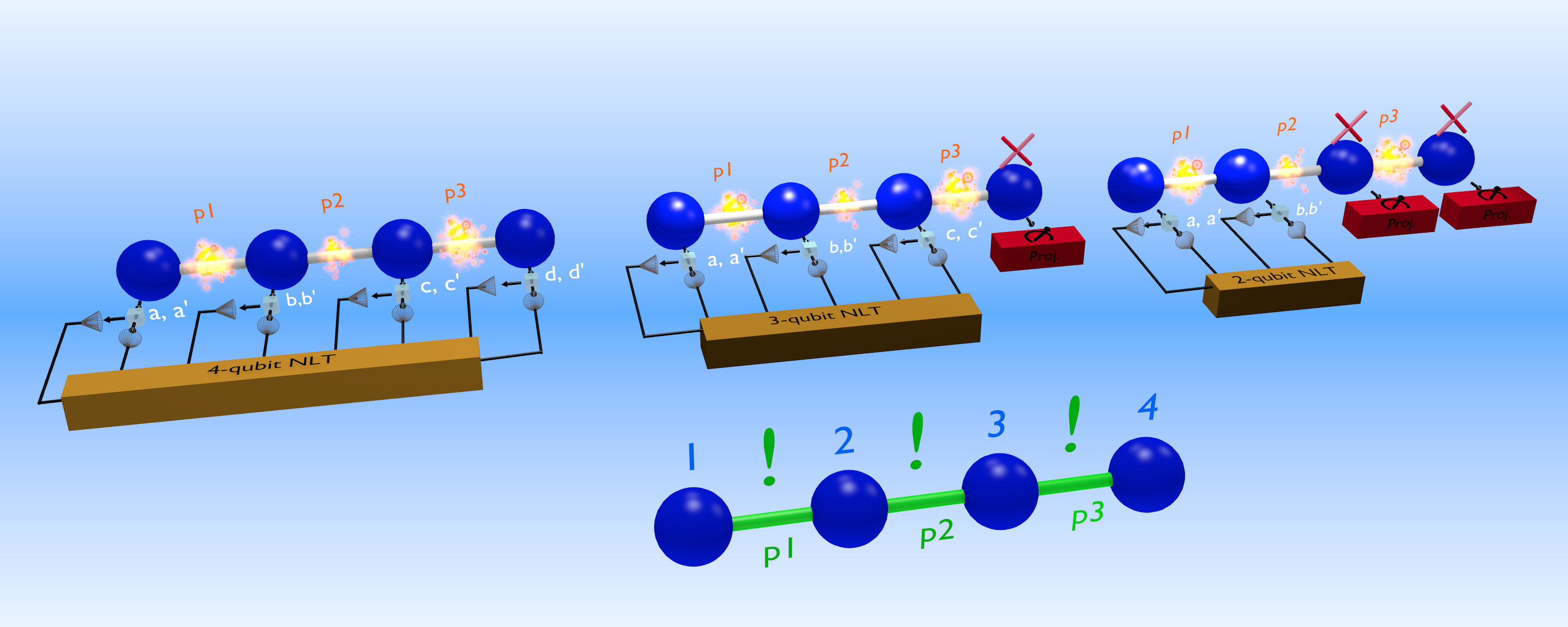}
\caption{(Color online) {\bf Panel (a):} A cluster state with given topology needs being analysed. This amounts to assigning a number to each link that describes concisely how well that connection is established. {\bf Panel (b):} The strategy for nonlocality-based network diagnostics: a set of multipartite nonlocality test is conducted on the whole cluster and to chosen subsets. From the results of these test, we can obtain a quantitative estimation of how well the connections are effected.}
\label{fig:clusterignoto}
\end{figure*}
%%%%%%%%%%%%%%%%%

With the necessity of efficiently manipulating quantum networks of increasing complexity comes the demand for reliable methods to implement the effective {\it diagnosis} of possible imperfections at both the preparation and operative stages. In turn, information about the quality of the operations that are used to synthesize a network will be invaluable for the design of better construction stages. Recently, various strategies based on statistical inference applied to quantum walk-like dynamics have been identified for the tracking of the faulty behaviour of a node, or a bond of one of such networks~\cite{references}. Needless to say, the elements of a network might not just share a physical link, but could be connected via quantum correlations. In this case, the interest of engineered diagnostic strategies would be that of ascertaining the structure and quality of the shared quantum correlations. Information gathered in this respect will be key to the design of better non-classical resources. This will be even more important in quantum information processing paradigms such as the measurement-based one~\cite{Briegel}, where the availability of specifically crafted entangled resources, cluster or graph states, is crucial to the success of any computational task. The relevance of such resources has prompted several experimental realisations~\cite{Walther05, Vallone07, Ceccarelli09, Su12, Monz11, Lanyon13, Yokoyama13, Chen14}, some of which have highlighted their networking potential~\cite{Aoki09,Armstrong12, Roslund14}.

%Within this view, the same faults as for standard quantum networks are still possible: failures of either the individual nodes or communication links - {\it i.e.} quantum correlations - due to the presence of some form of noise, peculiar to the physical platform. 

In this Letter we propose an approach that is radically different from any previous one for network diagnostics considered so far~\cite{references}. We build our strategy on the information about the structure of quantum correlations provided by the assessment of multipartite nonlocality inequalities (MNLIs). The rationale behind our approach is that, by post-processing the data provided by MNLIs,  useful information on both two-qubit entangling operations and single-qubit preparation stages can be gathered, even in those cases where an assessment based on the direct quantification of entanglement would be problematic, such as in the presence of multipartite mixed resources. In particular, we show that quantitative bounds to the quality of individual nodes and bonds of the assessed network can be established through our method. We demonstrate the effectiveness of the proposed approach by addressing experimentally a two-photon, four-qubit cluster state and using, as a quantitative instrument, the inequality proposed by Werner and Wolf ~\cite{Werner02} and, independently, \. Zukowski and Brukner~\cite{Zukowski02}, which we dub WWZB. The proposed tool incorporates a sufficient degree of flexibility to be insightful without the complications entailed by a test for genuine multipartite nonlocality. A violation of this inequaility quantifies through our simple model the strength of the links between the qubits of an addressed experimental resource. Moreover it enables a more powerful diagnosis than the simple assessment of two-qubit nonlocality tests, as it is able to address {\it generalized bipartitions}, thus attacking directly the implications of the sharing of quantum correlations above and beyond any study on two-qubit quantum correlations. 

We show that our diagnostic tool is informative enough to bound the amount of local noise acting on individual qubits of the network. When combined with pre-available knowledge on the features of a given network to be tested, our tool allows for the localisation of the source of single-qubit noise. %However, the actual localisation of the source of single-qubit noise remains somehow ambiguous when using only the tool embodied by our proposal.

\noindent
{\it The diagnostic tool.} The situation we address is illustrated in Fig.~\ref{fig:clusterignoto} {\bf a}. We consider a network of generally interconnected qubits, whose quality we would like to characterise. The connections among the network elements could be embodied by either physical interaction channels or general quantum correlated ones, such as in the situation that is explicitly illustrated here. While we assume to have full knowledge of the {shape} of the network (i.e. we assume knowledge of the adjacency matrix of the network), we do not know {\it how well} the nodes are actually connected. In this sense, the problem of assessing the quality of the state is reduced to that of assigning a quality measure to each link. 

As anticipated above, the quantitative figure of merit that we deploy to the diagnosis of the quality of our network is primarily embodied by the WWBZ inequality~\cite{Werner02,Zukowski02}, which we now briefly introduce for the sake of a self-contained presentation. 

Consider $N$ agents, each endowed with the possibility to choose between two dichotomic observables $\{\hat A_j({\bm n}_1),\hat A_j({\bm n}_2)\}$ ($j=1\dots N$), where ${\bf n}_k$ are local vectors in the single-qubit Bloch sphere, and which have been rescaled so that they can only take values $\pm 1$. For local realistic theories, the correlation function for the choice of local observables is thus $E(\{k_j\})=\langle\otimes^N_{j=1}\hat A_j({\bf n}_{k_j})\rangle$ ($k_j=1,2$). By choosing a suitable function $S(\{s_j\})$ that can take, again, only values $\pm1$ and depends on the indices $s_j\in\{-1,1\}$, one can derive the following family of $4^N$ Bell inequalities~\cite{Zukowski02} 
\begin{equation}
\label{Zuk}
\left\vert\sum_{\{s_j\}=\pm1}S(\{s_j\})\sum_{\{k_j\}=1,2}\left(\prod^N_{j=1} s^{k_j-1}_{j}\right)E(\{k_j\})\right\vert\le 2^N,
\end{equation}
whose right-hand side holds for local realistic theories. Eq.~\eqref{Zuk} contains interesting instances of Bell inequalities for $N$ particles, being trivially identical to the Clauser-Horne-Shimony-Holt version of Bell's inequality for $N=2$~\cite{CHSH}. It is possible to show that the fulfilment of Eq.~\eqref{Zuk} implies the possibility to construct local realistic models for the correlation function $E(\{k_j\})$, thus establishing such a family of inequalities as necessary and sufficient conditions for the local realistic description of the correlation function of an $N$-partite system~\cite{Zukowski02}.

In what follows, we make the choice of $S(\{s_j\})=\sqrt2\cos[\pi/4(\sum_js_j-N-1)]$, which allows us to recover the Mermin-Ardehali-Belinskii-Klyshko (MABK) inequality. Eq.~\eqref{Zuk} embodies the main tool for the diagnostic study that is at the core of this work. 
%As it will be seen through the analysis presented in the remainder of this work, Eq.~\eqref{Zuk} provides enough flexibility to assess the quality of a network of %connected particles without the need for the assessment of demanding figures of merit that probe the {\it genuinely} multipartite features of the network itself. %Yet, our proposed tool is more powerful than a simple analysis of, say, two-particle Bell inequalities, as it inherently relies on multi-particle correlation functions, %thus probing the features of the network at a higher level of complexity. 
In order to assess the features of our proposal, we specialise our study to the case of a linear cluster state, as illustrated in Fig.~\ref{fig:clusterignoto} {\bf b}. As it will be made evident, such an example is significant and motivated, as it addresses a network of correlated information carriers correlated in a genuinely multipartite fashion. 

\begin{figure*}[t]
\includegraphics[width=0.9 \linewidth]{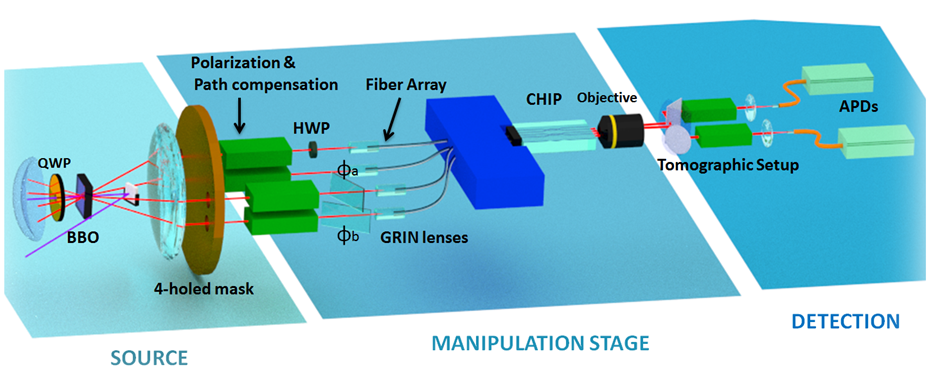}
\caption{(Color online) The experimental setup consists of a path-polarization hyperentangled source that generates the state $|\Xi \rangle =\frac{1}{2}(|HH\rangle_{AB}+|VV\rangle_{AB})\otimes (|\ell r\rangle_{AB}+|r \ell\rangle_{AB})$ \cite{Cinelli05,Vallone07}; the source is based on the use of a 1.5-mm Beta-Barium borate (BBO) crystal within an interferometric scheme, pumped with a 100 mW laser at $\lambda_p{=}355$nm. Degenerate photons are produced over a filter bandwidth of 6nm, and coupled in single mode fibres, delivering them to a femtosecond-laser written chip~\cite{Ciampini16}. This requires suitable polarisation compensation of the action of the fibres on the polarisation; further, a half waveplate (HWP) is put on the $l_a$ mode in order to generate a linear cluster state~ by performing a C-Phase operation between polarisation and path of the same photon~\cite{Vallone07}. The chip hosts two beam-splitters that are used, in a combination with the phase retarders
$\phi_A$ and $\phi_B$ to change the basis of the path qubits; polarisation analysis is performed by a standard tomographic setup. Results are obtained by measuring coincidence counts over two of the four output modes using single photon detectors. The typical counting rate through the chip was 50 coincidences/s.}
\label{fig:clustersetup}
\end{figure*}

\noindent
{\it The experimental study.} The experimental demonstration of the effectiveness of our proposal makes use of the resource embodied by a two-photon four-qubit cluster state engineered by means of the hyperentangled platform shown in Fig.~\ref{fig:clustersetup}. 

 %%%%%% Figure 2 %%%%%%
%%%%%%%%%%%%%%%%%

A strategy for obtaining an appropriate metric is suggested by the standard procedure for building arbitrary clusters: first each qubit is initialised in the superposition of its logical states $\ket{+}{=}\left({\ket{0}+\ket{1}}\right)/\sqrt{2}$; next, a controlled-Phase (C-Phase) gate is applied to each pair of nodes that need being linked. The quality of the link can then be traced back to the quality of the C-Phase gate that has been used. We have formalised the connection between the results of WWZB nonlocality tests on a cluster and a measurement of the fidelity of the underlying gates. The tests are conducted on the whole cluster, as well as subsections in which qubits are excluded by means of a measurement. By this connection, we obtain a number assessing the quality of each link from the experimental values of WWZB inequalities (Fig.1b). It is important to stress that the diagnostic strategy proposed here addresses the quality of a given resource, not the actual implementation strategy chosen to accomplish this task. Therefore, our methodology can be applied {\it tout court} to any other resource, regardless of its implementation. 

Further, in some architectures, entangling operations are implemented with high fidelity, while the state of the nodes can be corrupted by noise processes, peculiar to the physical system. For instance, dissipation mechanisms (including amplitude damping), should be taken into account in atomic or atom-like systems. In photonics, the loss of quantum entanglement can be usually described in terms of pure dephasing. We have investigated the possibility of pursuing our approach in the presence of perfect gates and noisy qubits.

We illustrate our method in a photonic implementation, in which we realise a four-qubit linear cluster states by two-photon hyperentanglement~\cite{Vallone07, Ciampini16}, adopting the setup shown in Fig.~\ref{fig:clustersetup}. This can produce a linear cluster in the form:
\begin{equation}
|C_4\rangle =\frac{1}{2}(\ket{H_aH_br_al_b}+\ket{V_aV_br_al_b}+\ket{H_aH_bl_ar_b}-\ket{V_aV_bl_ar_b}),
\end{equation}
where $H_x$ ($V_x$) denotes the horizontal (vertical) polarisation of the photon $x=a,b$, while $r_x$ ($l_x$) denotes a photon taking the right (left) path. We have performed a measurement of the four-party WWZB correlators: we have observed an experimental value of $18.53\pm0.23$, which has to be compared with the local realistic limit $2^4=16$, and with the quantum expectation $16\sqrt{2}\simeq22.63$. The clear deviation of the actual value from the ideal prediction flags the presence of reduced correlations within the cluster network. For the complete analysis, we have then measured WWZB correlators for different sub-partitions of the cluster, obtained by excluding the unwanted qubits by a suitable measurement; our results are reported in Table I. Our task is then to account for the observed violations of all the WWZB inequalities for all the eleven possible four-, three- and two-qubit groupings by comparing the actual results 
with those of a theoretical specular resource corrupted by noise. The amount of noise that reproduces the values obtained will be a measure of the quality of the cluster realised in the laboratory. 
  
%\bigskip{}
%
%\begin{center}
\begin{table}[h]
 \begin{tabular}{c c c}
 	\hline
 	\hline\textbf{qubit group} & {$WWZB_{max}$}  & {$WWZB_{exp}$}  \\ 
 	\hline $1-2-4 \equiv (\pi_A, \pi_B, k_B) $ & 11.31 & $9.32\pm0.19$  \\
 	\hline $1-2-3 \equiv (\pi_A,\pi_B, k_A)$ & 11.31 & $9.25\pm0.19$  \\
 	\hline $1-3-4 \equiv (\pi_A,k_A , k_B)$ & 13.66 & $11.71\pm0.17$ \\
 	\hline $2-3-4 \equiv (\pi_B, k_A,  k_B)$ & 13.66&  $11.08\pm0.13$  \\
 	\hline
      \hline
 	\hline $1-4 \equiv (\pi_A - k_B) $ & 5.66 & $4.55\pm0.13$ \\
 	\hline $1-3\equiv (\pi_A - k_A)$ & 5.66  & $4.62\pm0.13$ \\
 	\hline $2-3\equiv (\pi_B - k_A)$ & 5.66  & $4.33\pm0.15$  \\
 	\hline $2-4\equiv (\pi_B - k_B)$ & 5.66  &  $4.69\pm0.17$ \\
 	\hline $1-2\equiv (\pi_A - \pi_B)$ & 5.66  & $4.97\pm0.14$ \\
 	\hline $3-4\equiv (k_A - k_B)$ & 5.66 &  $4.50\pm0.14$ \\
 	\hline
	\hline
 \end{tabular} 
 \label{tableI}
\caption{Summary of the observed violations of the WWZB inequality for different qubit grouping within the cluster.}
\end{table}

\noindent
{\it Noise modelling.} Our four-qubit cluster state can in principle be obtained by applying a chain of three C-Phase gates to an initial $\vert{++++}\rangle_{1234}$ state of four separable qubits; we first model the nonideal behaviour of the gates by allowing for a failure probability $1{-}p$; the operation of the gate will then be described by a Kraus map of the form: \\
$M(\hat{\rho}) = p\; (\hat{U}_{\rm cp}\, \hat\rho \, \hat{U}^\dagger_{\rm cp}) +  (1-p) \hat\rho$, where $\hat{U}$ is the C-Phase operation. We then apply three maps on the initial state, realising the chain: $\hat{\rho}_{fin}(p_1, p_2, p_3)=M_{34}( M_{23}( M_{12}(\hat{\rho}_{in})))$. In this way, we can express the linear cluster state as a function of the probabilities that describe the C-Phase entangling gates; in turn, this gives expressions for the eleven WWZB parameters as a function of $(p_1, p_2, p_3)$. We can then find the values $(p_1^*, p_2^*, p_3^* )$ that best describe the actual violations, by minimising the distance of the predictions to the observations: $ (p_1^*,p_2^*,p_3^*)=\text{argmin}\;\sum_{i=1}^{11}\vert WWZB_i(p_1, p_2, p_3) - WWZB_i^{exp}\vert,$ where the summation index runs over the 11 subgroupings. The reliability of this optimisation has been tested numerically on a simulated corrupted cluster (see Supplementary Information). 

\begin{figure}[t]
\begin{centering}
\includegraphics[scale=0.045]{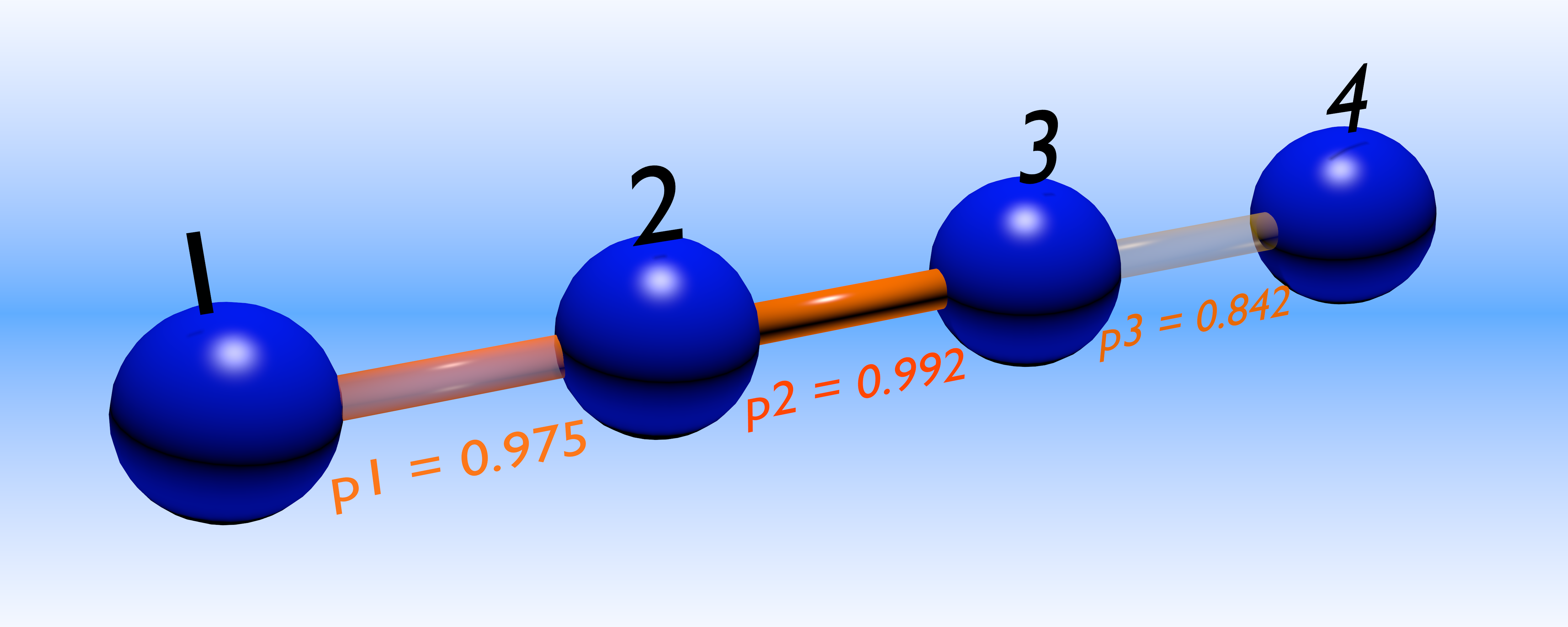}
\par\end{centering}
\caption{Link strength for a 4-qubit linear cluster state, using faulty-gates, each succeding with probability $p_i$. The problem of assessing the quality of the state is reduced to that of assigning a quality measure to each link. $p_i$ can assume values ranging from 0 to 1: $p_i=0$ implies full failure of the C-Phase operation in the building process of the cluster state, while $p_i=1$ implies its full success.}
\label{fig:finale}
\end{figure}

The analysis gives the results $ p_1^*= 0.975\pm0.024$, $p_2^*=0.992 \pm0.010$,  $p_3^*= 0.842\pm0.022$ (see Fig. \ref{fig:finale}). These values give the indication that the weakest link between the qubits is the one connecting the two path qubits, stemming from a reduced quality of the corresponding entangled resource. This observation is supported by direct experimental inspection, and it is likely due to unavoidable spatial phase instabilities present in our experimental scheme. 

It could be argued that in quantum photonics systems, failures of real-world gates are seldom described by our modelling; a commonplace imperfection is rather the loss of coherence, as described by single-qubit dephasing channels in the form $\varepsilon(\hat\rho){=}p\hat\rho+(1-p)\hat\sigma_z\hat\rho\hat\sigma_z$ ($\hat\sigma_z$ is the third Pauli matrix). We can repeat our analysis by adopting such a different noise model, and consider four dephasing channels acting on the cluster qubits, each with its own probability $p_1,p_2,p_3$, and $p_4$, along with perfect C-Phase gates. Direct inspection reveals that the predicted WWZB correlators only depend on the products $p_1p_2$ and $p_3p_4$. This is expected since, in this specific case, a dephasing channel on the first qubit can be replaced with an equivalent one acting on the second, obtaining the same theoretical expressions, and likewise for the fourth and third. The values we obtained are: $p_1^*p_2^*=0.913\pm0.051$, and $p_3^*p_4^*=0.892\pm0.060$. These can be somehow interpreted as an effective strength of the nodes 2 and 3 - rather than of the links - and these values too support the previous diagnosis that path entanglement is primarily responsible for the imperfections in the whole cluster state~\cite{nota1}.

Universal models are handy, but, due to their generality, they represent a Hegelian Night~\cite{Hegel}. With minimal inspection of the physics governing the generation of our cluster, we can obtain a more refined model. As a first example, we can observe that, while the initial polarisation and path entangled states are directly produced by our source, the final cluster is obtained by implementing a C-Phase gate between polarisation and path degrees of freedom of the same photon. 
As seen in Fig. 2 the cluster state $|C_4\rangle$ is experimentally engineered by introducing a half-waveplate at zero degrees over mode $\ell_A$. Starting from state $|\Xi\rangle$ , this produces the following transformation over photon A: $|H\ell\rangle_A \rightarrow |H\ell\rangle_A$, $|Hr\rangle_A \rightarrow |Hr\rangle_A$, $|Vr\rangle_A \rightarrow |Vr\rangle_A$, $|V\ell\rangle_A \rightarrow -|V\ell\rangle_A$. This represents a C-Phase operation between the target polarization qubit and the control path qubit of photon A.

We can use the depolarisation to describe the corruption of the gate between qubits 1-2 ($p_1$)and 3-4 ($p_3$), and use the probabilistic model $M(\hat\rho)$ ($p_2$) in order to describe the gate between qubits 2-3. This analysis gives the values $p_1^*=0.909\pm 0.019$, and $p_3^*= 0.901\pm 0.017$ for the action of the dephasing, and $p_2^*=0.980\pm 0.012$, once again in qualitative agreement with simpler models. This approach can be extended by including further depolarisation (captured by a probability $p_g$) acting identically on every qubit as a result of traversing the chip; in this case we get: 
$p_1^*= 0.986\pm 0.025, p_2^*=0.996\pm 0.007 , p_3^*= 0.967\pm 0.043$, and $p_g^*=0.866\pm 0.056$.

%$p_1^*= 0.980\pm??, p_2^*=1.00 \pm??, p_3^*= 0.982\pm??$, and $p_g^*=0.851\pm??$.

\noindent
{\it Conclusions and outlook.} We have experimentally assessed a diagnostic method able to probe the quality of bonds and links in a complex network of correlated particles. The methodology that we propose, albeit demonstrated explicitly on a specific instance of multipartite entangled state of qubits, is applicable to arbitrarily connected networks of information carriers, and makes no assumptions on the form of noise affecting its connections. As any diagnostics technique based on modelling, our approach is feasible for medium-sized networks comprising a few tens of qubits. However, due to the intrinsically modular nature of cluster states, our approach is still pertinent to section-by-section analysis of large photonics resources cluster, such as those built 'just in time' proposed in Ref.~\cite{Benjamin}.

\noindent
{\it Acknowledgements.--} MP acknowledges support from the Collaborative Project TherMiQ, the John Templeton Foundation (grant number 43467), the Julian Schwinger Foundation (grant number JSF-14-7-0000), and the UK EPSRC (grant number EP/M003019/1). MB was supported by MIUR with a Rita Levi-Montalcini fellowship. We acknowledge financial support by the H2020-FETPROACT-2014 Grant QUCHIP (grant agreement 641039). FS is supported by the European Research Council through the Starting Grant 3D-QUEST (grant agreement 307783).


\begin{thebibliography}{99}
\bibitem{nota1} Such phase instabilities are likely to be due to the presence of the finer array that couples the free-space radiation to the chip.
\end{thebibliography}
\end{document}